\documentclass[
aps,pra,
twocolumn
]{revtex4}
\usepackage{graphicx}
\usepackage{subfigure}

\usepackage{amsmath}
\usepackage{amssymb}
\usepackage{amsthm}
\usepackage{braket}
\usepackage{color}
\usepackage{microtype}
\usepackage{tikz}
\usepackage{makecell}

\usepackage{hyperref}
\hypersetup{colorlinks,linkcolor={blue},citecolor={red},urlcolor={blue}}

\theoremstyle{plain}

\theoremstyle{plain}

\begin{document}

\title{Mass transfer and boson cloud depletion in a binary black hole system}

\author{Yao Guo}
\author{Wenjie Zhong}
\author{Yiqiu Ma}
\email{myqphy@hust.edu.cn}
\author{Daiqin Su}
\email{sudaiqin@hust.edu.cn}

\affiliation{ MOE Key Laboratory of Fundamental Physical Quantities Measurement, Hubei Key Laboratory of Gravitation and Quantum Physics, PGMF, Institute for Quantum Science and Engineering, School of Physics, Huazhong University of Science and Technology, Wuhan 430074, China}


\date{\today}

\begin{abstract}
Ultralight boson is one of the potential candidates for dark matter. If exists, it can be generated by a rapidly rotating black hole via superradiance, extracting the energy and angular momentum of the black hole and forming a boson cloud. The boson cloud can be affected by the presence of a companion star, generating fruitful dynamical effects and producing characteristic gravitational wave signals. We study the dynamics of the boson cloud in a binary black hole system, in particular, we develop a framework to study the mass transfer between two black holes. It is found that bosons occupying the growing modes of the central black hole can jump to the decaying modes of the companion black hole, resulting in cloud depletion. This mechanism of cloud depletion is different from that induced by the resonant perturbation from the companion.
\end{abstract}

\maketitle

\section{Introduction}

The detection of gravitational waves by current ground-based gravitational wave detectors~\cite{PhysRevLett.116.061102} (LIGO, Virgo, etc.) opens up a new avenue to explore the astrophysical processes that involve strong gravitational field. Future space-borne gravitational wave detectors (LISA, TQ~\cite{luo2016tianqin}, etc.) extend the detection frequency range and allow for exploration of more fruitful physics, e.g., supermassive black holes. One of the interesting target sources for gravitational wave detectors is the ultralight boson, including axion and pseudoscalar axion-like particles. Axion provides a solution to the strong CP problem~\cite{PhysRevLett.38.1440, PhysRevLett.40.223, PhysRevLett.40.279} and has the potential to demystifying the baryon asymmetry of the universe, while more general axion-like particles are predicted by symmetry breaking in string theory~\cite{PeterSvrcek_2006}. These ultralight bosons are also potential candidates for dark matter~\cite{PRESKILL1983127, ABBOTT1983133, DINE1983137, PhysRevD.89.083536, PhysRevD.95.043541}. The ultralight bosons, if exists, can be produced by a rapidly rotating black hole (BH) through superradiance instabilities~\cite{ZelDovich1971generation, ZelDovich1972Amplification, Misner1972Stability, Starobinski1973Amplification, Starobinski1974Amplification, PhysRevLett.119.041101, PhysRevD.96.024004, brito2020superradiance,Cromb_2020}, carrying away the mass and angular momentum of black hole and forming a boson cloud/condensate~\cite{PhysRevD.83.044026}. 

Since the presence of ultralight boson reduces the mass and spin of the black hole, the precision measurement of black hole mass and spin via gravitational wave detectors provides a powerful constraint on the properties of bosons~\cite{PhysRevD.96.064050,Arvanitaki_2015,Arvanitaki_2010}. In addition, the boson cloud can radiate continuous gravitational waves due to its asymmetric distribution and self-annihilation~\cite{PhysRevD.83.044026}, and thus can be detected by future gravitational wave detectors. 

If a companion star is present, the boson cloud around a black hole is distorted by a time-dependent tidal field. Under certain conditions, the time-dependent tidal field can induce transitions between the boson's various energy levels, in particular, between the growing modes and decaying modes~\cite{PhysRevD.99.044001}. This would generate fruitful dynamical effects. The transition of boson can transfer energy and angular momentum from the boson cloud to the companion star, modifying its orbital evolution. In the extreme case, the energy loss of the companion star due to the emission of gravitational waves is balanced by the energy gain from the boson cloud, thus forming the so called ``floating orbit"~\cite{PhysRevD.99.064018}. When the boson occupies the decaying mode, it may decay into the black hole, resulting in the depletion of the boson cloud~\cite{PhysRevD.99.044001, PhysRevD.99.104039}. In some cases, the boson cloud could be completely cleaned up by the companion star. The bosons that absorbed by the black hole increase the mass of the black hole while reduce its spin. The reduction of the black hole spin can turn some of the growing modes into decaying modes and then accelerate the cloud depletion~\cite{PhysRevD.107.103020}. The companion can also induce transitions between bound and unbound orbits of the boson, thus ionize it~\cite{PhysRevLett.128.221102}. These fruitful dynamical processes of the boson cloud produce characteristic gravitational wave signals and consequently can be detected by current and future gravitational wave detectors~\cite{PhysRevD.107.104003}. This provides a way to infer the presence of the ultralight boson and constrain its various properties. 

When the black hole and its companion is sufficiently close, the boson can escape to the companion, resulting in a redistribution of the mass of the boson cloud. A gravitational ``molecule", an analog to the hydrogen molecule, can be formed under certain conditions~\cite{PhysRevD.101.124049, PhysRevD.103.024020}. 
If the companion is also a rotating black hole, then the escaped boson may occupy the decaying modes of the companion black hole and may decay into it. If happens, then this provides a new channel for boson cloud depletion. In this work, we develop a framework to study the mass transfer between two black holes in a binary system, assuming that the two black holes have the same mass and spin, and their spin orientation is parallel. We make analogy with the hydrogen molecule ion and calculate the wave functions of the molecular orbits and the corresponding energy eigenvalues using the variational method. We then obtain the probability for the boson jumping to the decaying mode of the companion black hole, and show that this leads to a strong cloud depletion which almost completely cleans up the boson cloud. 

This paper is organized as follows. In Sec.~\ref{sec:background}, we briefly review the boson cloud around a black hole and its depletion into the black hole due to the time-dependent tidal field produced by the companion. In Sec.~\ref{sec:orbit}, by making analogy with hydrogen molecular ion, we use the variational method to derive the wave functions and energy eigenvalues of the boson in the binary black hole system. In Sec.~\ref{sec:depletion}, we use the adiabatic approximation to study the time evolution of the wave function of the boson, evaluate the the probability of jumping to the decaying mode of the companion black hole, and calculate the time evolution of the cloud mass due to the decay into the companion black hole. We finally conclude in Sec.~\ref{sec:conclusion}. 

\section{Boson cloud around a black hole}\label{sec:background}


\subsection{Gravitational atom}

A rapidly rotating black hole can radiate ultralight bosons via superradiance instabilities. These bosons condensate in some of their orbits, forming a boson cloud around the black hole. When the mass of the boson is small, the size of the cloud can be much larger than the gravitational radius of the rotating black hole. In this limit, the Newtonian approximation is sufficient to describe the dynamics of the boson cloud. The orbits of the boson is determined by a Schr\"odinger-like equation, similar to that of an electron in a hydrogen atom. The eigenstates of the boson are denoted as $\ket{\varphi_{n \ell m}}$ or $\varphi_{n \ell m}$, and the eigenfrequencies are given by~\cite{PhysRevD.99.044001}
\begin{eqnarray}
\omega_{n \ell m} \approx \mu \bigg( 1 - \frac{\alpha^2}{2 n^2} \bigg),
\end{eqnarray}
where $\mu$ is the mass of the boson, $\alpha \equiv G M \mu/\hbar c$ is the dimensionless ``fine-structure constant". The radial profile of the wave function peaks at  
\begin{eqnarray}
r_{c, n} \approx \bigg( \frac{n^2}{\alpha^2} \bigg) r_g = n^2 r_{b}, 
\end{eqnarray}
where $r_g$ is defined as the gravitational radius of the black hole, $r_g \equiv GM/c^2$, and $r_{b} = r_g/\alpha^2$ is defined as the Bohr radius.  

However, there is a crucial difference between the electron in the hydrogen atom and the boson around a black hole: the orbits of the electron are stable while the orbits of the boson are not stable due to the presence of the black hole horizon. This is characterized by the imaginary part of the eigenfrequency, $\omega_{n \ell m} \rightarrow \omega_{n \ell m} + i \Gamma_{n \ell m}$. In the limit $\alpha  \ll 1$, $\Gamma_{n \ell m}$ can be approximated as~\cite{PhysRevD.22.2323}
\begin{eqnarray}\label{eq:decayrate}
\Gamma_{n \ell m} = \frac{2 r_+}{M} C_{n \ell m}(\alpha) (m \Omega_{\rm H} - \omega_{n \ell m}) \alpha^{4\ell +5},
\end{eqnarray}
where $C_{n \ell m}(\alpha)$ is positive and given by
\begin{eqnarray}
C_{n \ell m}(\alpha) &=& \frac{2^{4\ell+1} (n+\ell)!}{n^{2\ell+4}(n-\ell-1)!} \bigg[ \frac{\ell !}{(2\ell)! (2\ell+1)!} \bigg]^2 \nonumber \\
&& \times \prod_{j=1}^\ell \bigg[ j^2(1-\tilde{a}^2) + (\tilde{a} m - 2 \tilde{r}_+ \alpha)^2 \bigg],
\end{eqnarray}
with $\tilde{a} = a/M$ and $\tilde{r}_+ = r_+/M$. Here $r_+$ is the size of the event horizon, $a$ is the spin and $\Omega_{\rm H}$ is the angular velocity of the rotating black hole. The orbits with positive $\Gamma_{n \ell m}$ are growing modes, for which the number of boson grows exponentially; while the orbits with negative $\Gamma_{n \ell m}$ are decaying modes, for which the number of boson decays exponentially. Starting from a rapidly rotating black hole, bosons are radiated due to the superradiance instabilities and then occupy the growing modes. The radiated bosons carry away angular momentum, slowing down the rotation of the black hole. At an equilibrium point the black hole rotates slow enough such that no bosons can be further radiated, and a quasi-stationary boson cloud forms around the black hole. 

\subsection{Hyperfine and Bohr resonance}

When the black hole with a boson cloud is part of a binary system, the gravitational field of the companion star distorts the cloud, resulting in transition of bosons between growing modes and decaying modes. The bosons that jump to decaying modes can return to the black hole, reducing the total mass of the cloud and transferring angular momentum to the companion star. Therefore,  the existence of boson cloud would affect the orbital evolution of the companion star and the gravitational waveforms from the binary system.  

The companion star induces a time-dependent perturbation to the Kerr metric, which then introduces a time-dependent shift of gravitational potential to the Schr\"odinger equation that dominates the dynamics of bosons. Under certain conditions, the time-dependent perturbation can induce resonant transitions between growing modes and decaying modes~\cite{PhysRevD.99.044001}. There are two types of resonances, the hyperfine (or Rabi) resonance and the Bohr resonance. The resonance occurs at a specific orbital separation. For the hyperfine resonance, the orbital separation is given by
\begin{eqnarray}
R_{\rm res}^{(h)} = 144^{1/3} \alpha^{-4} (1+q)^{1/3} \tilde{a}^{-2/3} r_g,
\end{eqnarray}
while for the Bohr resonance, the orbital separation is given by  
\begin{eqnarray}
R_{\rm res}^{(b)} = \bigg( \frac{144}{5} \bigg)^{2/3} \alpha^{-2} (1+q)^{1/3}  r_g. 
\end{eqnarray}

In general, the orbital separation of the hyperfine resonance is much larger than that of the Bohr resonance. This is because the hyperfine energy gap is much narrower than the Bohr energy gap. This implies that the companion star has to be closer to the black hole in order to excite the Bohr resonance. 




\section{Boson orbits in a binary black hole system}\label{sec:orbit}

We are concerned with the dynamics of the boson cloud when a black hole and its companion are sufficiently close so that mass transfer between them cannot be ignored. The process of mass transfer is generally very complicated and numerical simulation is needed to fully uncover the cloud dynamics. It has been shown by using effective field theory techniques~\cite{PhysRevD.101.124049} and numerical calculation~\cite{PhysRevD.103.024020} that a gravitational molecule can form in a binary black hole system. Here, we use a simple model to describe the mass transfer and the time evolution of a BH-cloud-companion system. The model is based on the analogy between the hydrogen molecule ion $H_2^+$ and the BH-cloud-companion system, and it can capture the main characteristics of mass transfer and cloud depletion.

\begin{figure}[h]
\includegraphics[width=0.98 \columnwidth]{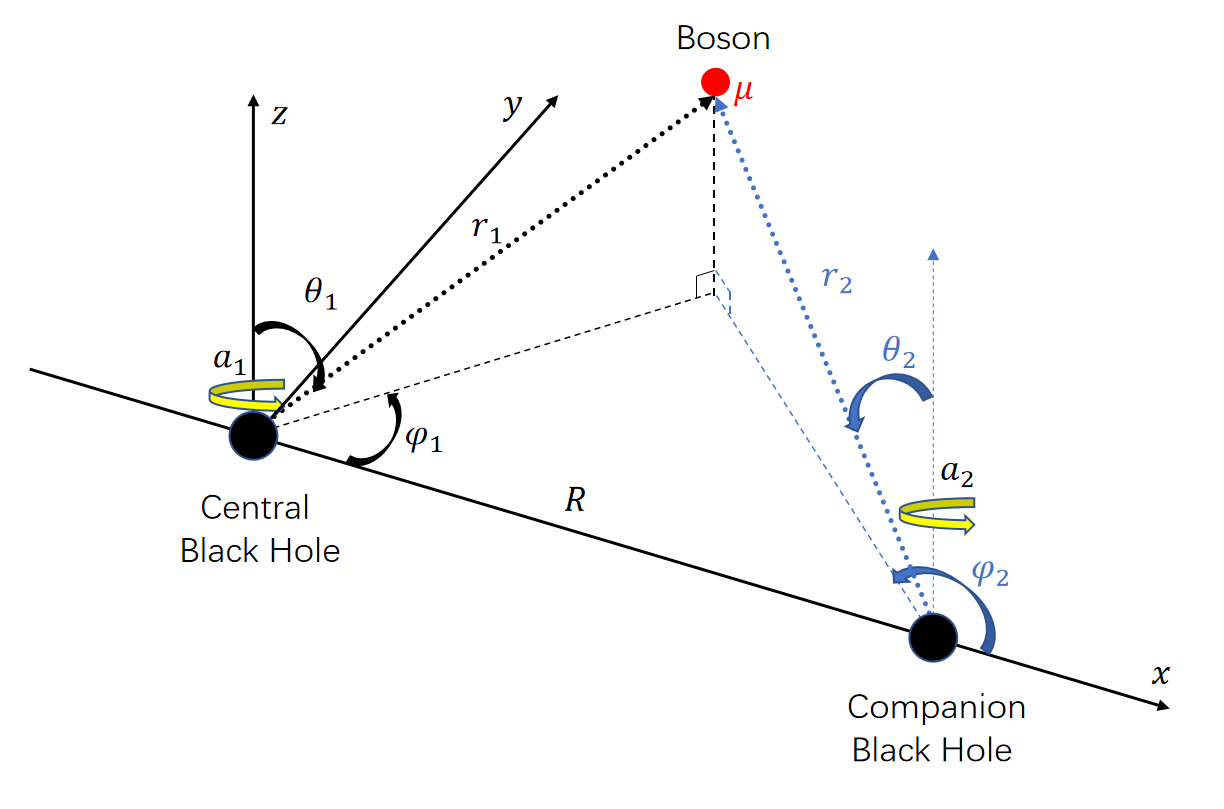}
\caption{ Configuration of the BH-cloud-BH system and the coordinate system used to describe the boson. The origin of the coordinate system is located at the central black hole, and the $z$-axis is parallel to the spin of the central black hole and the $x$-axis is pointing towards the companion black hole. Spherical coordinates $(r_1, \theta_1, \varphi_1)$ are used to represent the position of the boson relative to the central black hole, and $(r_2, \theta_2, \varphi_2)$ are used to denote the position of the boson relative to the companion black hole. Here $a_1$ and $a_2$ represent the spin of the central and companion black holes, respectively, and $R$ denote the orbital separation.}
\label{fig:configuration}
\end{figure}

For simplicity, we assume (1) the companion is also a black hole, therefore forming a binary black hole system; (2) two black holes have the same mass and spin; (3) their spin orientation is parallel and perpendicular to the orbital plane. The schematic of the configuration is shown in Fig.~\ref{fig:configuration}. These assumptions allow us to focus on the effect of mass transfer, and they can be relaxed to incorporate more complicated effects. Under these assumptions, the orbits of the boson in the BH-cloud-BH system is analogous to that of the electron in the hydrogen molecule ion $H_2^+$, except that the two black holes rotate with each other and that the existence of a horizon results in boson absorption. 

\subsection{Orbits of hydrogen molecule ion}\label{sec:orbits}

The hydrogen molecule ion $H_2^+$ consists of two protons and a single electron. The electron moves in the potential produced by the two protons with a fixed distance. The potential is time independent, so the electron wave functions can be derived by solving the stationary Schr\"odinger equation. Approximate energy eigenvalues and eigenfunctions can be solved using the variational method. 
The energy eigenvalues are the stationary points of the expectation value of the Hamiltonian. 
To approximate the energy eigenvalue, one starts from a trial wave function and then vary the parameters in the trial wave function to find the stationary point of the Hamiltonian. 

For the hydrogen molecule ion, the initial trial wave functions can be selected by using the symmetric properties of the system. For example, one can choose a linear superposition of the two ground-state wave functions as a trial function to find the ground state of the hydrogen molecule ion. The excited states can also be found in a similar way. In this paper, we are mainly interested in the excited states with $n = 2$ because these states are relevant to the states that dominate the evolution of the boson cloud. It can be shown that there exists two molecular $\sigma$ orbits obtained from the atomic orbit $2 p_x$~\cite{cohen1986quantum} (see Appendix~\ref{app:WaveFunc} for details), 
\begin{eqnarray}\label{eq:2px}
\ket{\sigma } &=& \frac{1}{\sqrt{2} N_1} \big( \ket{\varphi_{2p_x}^1} - \ket{\varphi_{2p_x}^2} \big) \nonumber\\
&=& \frac{1}{ 2 N_1} \big( \ket{\varphi_{2,1,1}^1} + \ket{\varphi_{2,1,-1}^1} - \ket{\varphi_{2,1,1}^2} - \ket{\varphi_{2,1,-1}^2} \big), \nonumber\\
\ket{\sigma^* } &=& \frac{1}{\sqrt{2} N_2} \big( \ket{\varphi_{2p_x}^1} + \ket{\varphi_{2p_x}^2} \big) \nonumber\\
&=& \frac{1}{ 2 N_2} \big( \ket{\varphi_{2,1,1}^1} + \ket{\varphi_{2,1,-1}^1} + \ket{\varphi_{2,1,1}^2} + \ket{\varphi_{2,1,-1}^2} \big); \nonumber\\
\end{eqnarray}
and another two molecular $\pi$ orbits obtained from the atomic orbit $2 p_y$,
\begin{eqnarray}\label{eq:2py}
\ket{\pi } &=& \frac{1}{\sqrt{2} N_3} \big( \ket{\varphi_{2p_y}^1} + \ket{\varphi_{2p_y}^2} \big) \nonumber\\
&=& \frac{1}{ 2 i N_3} \big( \ket{\varphi_{2,1,1}^1} - \ket{\varphi_{2,1,-1}^1} + \ket{\varphi_{2,1,1}^2} - \ket{\varphi_{2,1,-1}^2} \big), \nonumber\\
\ket{\pi^* } &=& \frac{1}{\sqrt{2} N_4} \big( \ket{\varphi_{2p_y}^1} - \ket{\varphi_{2p_y}^2} \big) \nonumber\\
&=& \frac{1}{ 2 i N_4} \big( \ket{\varphi_{2,1,1}^1} - \ket{\varphi_{2,1,-1}^1} - \ket{\varphi_{2,1,1}^2} + \ket{\varphi_{2,1,-1}^2} \big), \nonumber\\
\end{eqnarray}
where $N_1, N_2, N_3$ and $N_4$ are introduced to normalize these states,
\begin{eqnarray}\label{eq:Normalization}
N_1 &=& \sqrt{1 - \langle \varphi_{2p_x}^1 \ket{\varphi_{2p_x}^2}}, ~~~~  N_2 = \sqrt{1 + \langle \varphi_{2p_x}^1 \ket{\varphi_{2p_x}^2}}, \nonumber\\
N_3 &=& \sqrt{1 + \langle \varphi_{2p_y}^1 \ket{\varphi_{2p_y}^2}}, ~~~~ N_4 = \sqrt{1 - \langle \varphi_{2p_y}^1 \ket{\varphi_{2p_y}^2}}. \nonumber \\
\end{eqnarray}
It is evident that $N_i$ depends on the distance between two protons. In the following, we will use the notation $i = 1, 2, 3, 4$ to denote the four molecular orbits $\sigma, \sigma^*, \pi$ and $\pi^*$, respectively. 

\subsection{Born-Oppenheimer approximation}\label{section:BOapprox}

We now turn to the problem of solving the orbits of the boson in a binary black hole system with the two black holes have the same mass and spin. This system is analogous to the hydrogen molecule ion $H_2^+$ except that the two black holes rotate with each other and their separation shrinks due to the emission of gravitational waves. The bosons experience a time-dependent field rather than a static one. An important question is how to take into account the effects of rotation and orbital shrinking. 

In Ref.~\cite{PhysRevD.99.044001}, the rotation of the companion star relative to the central black hole produces a time-varying tidal field that perturbs the central black hole, which then induces hyperfine and Bohr mixing of growing and decaying modes, resulting in boson cloud depletion. Whilst we are concerned with the transfer of bosons between the two black holes, the framework developed in Ref.~\cite{PhysRevD.99.044001} does not apply here. As a first approximation, we neglect the effects of rotation and orbital shrinking, and treat the two black holes as stationary with a fixed orbital separation. This is known as the Born-Oppenheimer approximation, in which the binary black hole system with a boson cloud is analogous to the hydrogen molecule ion at any given time. 
Before performing the detailed calculation, we compare various timescales to validate the use of Born-Oppenheimer approximation. 

The velocity of a boson can be estimated using the Virial theorem and the uncertainty principle. The boson's velocity $v_a$ satisfies 
\begin{eqnarray}
\frac{1}{2} \mu v_a^2 \sim \frac{G M \mu}{r_c}, ~~~~~~
\mu v_a \sim \frac{\hbar}{r_c},
\end{eqnarray}
where $r_c$ is the typical length scale of the radial profile of the boson cloud. This implies 
\begin{eqnarray}
v_a \sim \frac{G M \mu}{\hbar c} c \sim \alpha c \sim \alpha.
\end{eqnarray}
The relaxation timescale for the boson can be approximated by
\begin{equation}
\tau_r\sim \frac{R}{v_{a}} \sim \frac{R}{\alpha}. 
\end{equation}
The time $\tau_r$ characterizes the timescale for the boson moving from one black hole to the other. The period of the binary black holes is simply given by
\begin{equation}
T=\frac{2\pi}{\Omega}=2\pi\sqrt{\frac{R^3}{M(1+q)}}.
\end{equation}
Therefore the ratio between $\tau_r$ and $T$ is 
\begin{eqnarray}
\frac{\tau_r}{T} \sim \frac{\sqrt{1+q}}{2 \pi} \sqrt{\frac{M}{\alpha^2 R}}
\sim \frac{\sqrt{1+q}}{2 \pi} \sqrt{\frac{r_b}{R}}.
\end{eqnarray}
For $q=1$ and 
$R = 32 \, r_b$, $\tau_r/T \sim 1/8\pi \sim 0.04$. As we will see in the following discussion, the mass transfer occurs when $R \gtrsim 50 \, r_b$. This shows that for an intermediate orbital separation, the relaxation time of the boson is much shorter than the period of the binary. Therefore, the two black holes can be treated as quasi-static when considering the transfer of bosons and the Born-Oppenheimer approximation applies. 

The orbit of the binary black holes shrinks due to the emission of gravitational waves~\cite{PhysRev.136.B1224},
\begin{eqnarray}\label{eq:R1}
R(t) = \bigg[ \frac{M(1+q)}{\Omega_0^2}\bigg]^{1/3} \bigg( -\frac{t}{\tau_0} \bigg)^{1/4},
\end{eqnarray}
where we set $t=0$ as the moment of merger, and $\tau_0$ is the time to merger for an initial orbital frequency $\Omega_0$. The initial time $\tau_0$ and initial orbital frequency $\Omega_0$ are related via 
\begin{eqnarray}\label{eq:R2}
\frac{\tau_0}{M(1+q)}=\frac{5}{256}\frac{(1+q)^2}{q}\bigg[ \frac{1}{M(1+q)\Omega_0} \bigg]^{8/3}.
\end{eqnarray}
The characteristic timescale for coalescence can be estimated via
\begin{equation}
\tau_p=\bigg|\bigg(\frac{d R}{d t}\bigg)^{-1} \bigg|R \approx \frac{5}{64}\frac{R^4}{M^3}\frac{1}{q(1+q)}. 
\end{equation}
This timescale is evidently much longer than the period of the binary $T$, and it is thus also much longer than the boson relaxation timescale $\tau_r$.
\begin{eqnarray}
\frac{\tau_r}{\tau_p} \sim \frac{32}{5} \alpha^5 \bigg(\frac{r_b}{R} \bigg)^3 q(1+q).
\end{eqnarray}
When the orbital separation of the two black holes is sufficiently large and $q$ is not so large, these two ratios are both much smaller than one, so the Born-Oppenheimer approximation is sound.


\subsection{Eigenfunctions and energy eigenvalues}\label{sec:eigen}

To find the exact eigenfunctions and eigenfrequencies, one in principle needs to solve the Klein-Gordon equation of the boson in the background spacetime of the binary black hole system, which is quite a challenging task. When the mass of the boson is small, the boson cloud is far away from the black hole and Newtonian approximation can be used to derive the eigenfunctions~\cite{PhysRevD.99.044001}. To the order of $1/r$, the wave function of the boson around a single rotating black hole satisfies 
\begin{eqnarray}
i \hbar \frac{\partial }{\partial t} \psi(t, \boldsymbol{r}) = \bigg( - \frac{1}{2 \mu} \nabla^2 - \frac{\alpha}{r} \bigg) \psi(t, \boldsymbol{r}), 
\end{eqnarray}
which is exactly in the same form as the Schr\"odinger equation for the electron in a hydrogen atom. When the two black holes are not so close to each other and the boson lingers around a regime far away from both black holes, then the Newtonian approximation applies. To the order of  $1/r$, the wave function of the boson in the binary black hole system satisfies
\begin{eqnarray}\label{eq:BoBinary}
i \hbar \frac{\partial }{\partial t} \psi(t, \boldsymbol{r}) = \bigg( - \frac{1}{2 \mu} \nabla^2 - \frac{\alpha}{r_1} - \frac{\alpha}{r_2} \bigg) \psi(t, \boldsymbol{r}), 
\end{eqnarray}
where $r_1$ is the distance between the boson and the central black hole, and $r_2$ is the distance between the boson and the companion black hole. Equation~\eqref{eq:BoBinary} is exactly in the same form as the Schr\"odinger equation for the electron in the hydrogen molecule ion, without including the interaction between two protons.

In the Born-Oppenheimer approximation, the eigenfunctions of the boson at any given time can be derived straightforwardly. They are exactly in the same form as those given by Eqs.~\eqref{eq:2px} and \eqref{eq:2py}, with $\ket{\varphi_{n \ell m}}$ the eigenfunctions of the boson in a single isolated rotating black hole. Furthermore, the normalization constants $N_i$ depend on the configuration of the BH-cloud-BH system. For the case that we consider here, the normalization constants have no analytic expressions and are needed to be evaluated numerically. 


%

Once the eigenfunctions are known, the energy eigenvalues can be calculated straightforwardly, which are simply the expectation values of the Hamiltonian. From Eq.~\eqref{eq:BoBinary} it is evident that the Hamiltonian of the boson in the Newtonian limit is 
\begin{eqnarray}
\hat{H} = - \frac{1}{2 \mu} \nabla^2 - \frac{\alpha}{r_1} - \frac{\alpha}{r_2}. 
\end{eqnarray}
To simplify the calculation, we divide the Hamiltonian $\hat{H}$ into two parts: the Hamiltonian of the boson in the central black hole and the potential produced by the companion black hole, namely, $\hat{H} = \hat{H}_1 - \frac{\alpha}{r_2}$ with $\hat{H}_1 = - \frac{1}{2 \mu} \nabla^2 - \frac{\alpha}{r_1}$. Then the eigenfrequencies are given by
\begin{widetext}
\begin{eqnarray}\label{eq:EigenFre}
\omega_1&=& \bra{\sigma} \hat{H} \ket{\sigma}
=\frac{1}{N_1^2} \bigg[-\frac{1}{8}\mu \alpha^2 \big(1 - \langle \varphi_{2p_x}^1 \ket{\varphi_{2p_x}^2} \big) - \bra{\varphi_{2p_x}^1} \frac{\alpha}{r_2} \ket{\varphi_{2p_x}^1} 
+ \bra{\varphi_{2p_x}^1} \frac{\alpha}{r_2} \ket{\varphi_{2p_x}^2} \bigg],  
\nonumber\\
\omega_2&=& \bra{\sigma^*} \hat{H} \ket{\sigma^*}
=\frac{1}{N_2^2} \bigg[-\frac{1}{8}\mu \alpha^2 \big(1 + \langle \varphi_{2p_x}^1 \ket{\varphi_{2p_x}^2} \big) - \bra{\varphi_{2p_x}^1} \frac{\alpha}{r_2} \ket{\varphi_{2p_x}^1} 
- \bra{\varphi_{2p_x}^1} \frac{\alpha}{r_2} \ket{\varphi_{2p_x}^2} \bigg],  
\nonumber\\
\omega_3&=& \bra{\pi} \hat{H} \ket{\pi}
=\frac{1}{N_3^2} \bigg[-\frac{1}{8}\mu \alpha^2 \big(1 + \langle \varphi_{2p_y}^1 \ket{\varphi_{2p_y}^2} \big) - \bra{\varphi_{2p_y}^1} \frac{\alpha}{r_2} \ket{\varphi_{2p_y}^1} 
- \bra{\varphi_{2p_y}^1} \frac{\alpha}{r_2} \ket{\varphi_{2p_y}^2} \bigg],  
\nonumber\\
\omega_4&=& \bra{\pi^*} \hat{H} \ket{\pi^*}
=\frac{1}{N_4^2} \bigg[-\frac{1}{8}\mu \alpha^2 \big(1 - \langle \varphi_{2p_y}^1 \ket{\varphi_{2p_y}^2} \big) - \bra{\varphi_{2p_y}^1} \frac{\alpha}{r_2} \ket{\varphi_{2p_y}^1} 
+ \bra{\varphi_{2p_y}^1} \frac{\alpha}{r_2} \ket{\varphi_{2p_y}^2} \bigg]. 
\end{eqnarray}
\end{widetext}
For the case that we consider here, there exists no analytic expressions for $\omega_i$, so they have to be evaluated numerically. Note that in deriving the eigenfunctions and eigenfrequencies, we have ignored the higher order corrections to the Hamiltonian. This results in degeneracy of all four energy levels when the two black holes are infinitely far away.

\section{Cloud depletion}\label{sec:depletion}

Assume that the central black hole has a boson cloud surrounding it while the companion black hole does not. When the two black holes are very far away from each other, the bosons move around the central black hole and cannot escape to the companion. When the two black holes become closer to each other, the boson orbits that belong to the central black hole and that belong to the companion have more overlap, forming boson orbits that are analogous to the molecular orbits of the hydrogen molecule ion. Therefore, the boson around the central black hole may jump to the companion. This can  have two consequences: first, the bosons redistribute in the binary black hole system, changing its quadruple and thus modifying the gravitational waveform; second, the bosons may jump to the decaying modes of the companion, and therefore could be absorbed by the companion black hole. To estimate the importance of the above two consequences, one needs to model the process of boson transfer from the central black hole to its companion.

The binary black hole system emits gravitational waves so its orbit shrinks, namely, the orbital separation between the two black holes shrinks due to the radiation of energy. The orbital separation, denoted as $R(t)$, is thus time dependent. This implies that the eigenfunctions and eigenfrequencies are also time dependent because they are solved at a given time assuming the orbital separation is fixed. These solutions are meaningful only in the case where the time scale of the orbital shrinking is much larger than that of the boson relaxation. This is exactly the case as we have discussed in Sec.~\ref{section:BOapprox}. In order to know the evolution of the boson cloud in the binary black hole system, one needs to solve a time-dependent Schr\"odinger equation, in which the Hamiltonian varies slowly. This sort of problem can be solved using the adiabatic approximation if the energy levels are not degenerate and the energy gap is sufficiently large. 

\subsection{Adiabatic approximation}

\begin{figure}[h]
\includegraphics[width=0.92 \columnwidth]{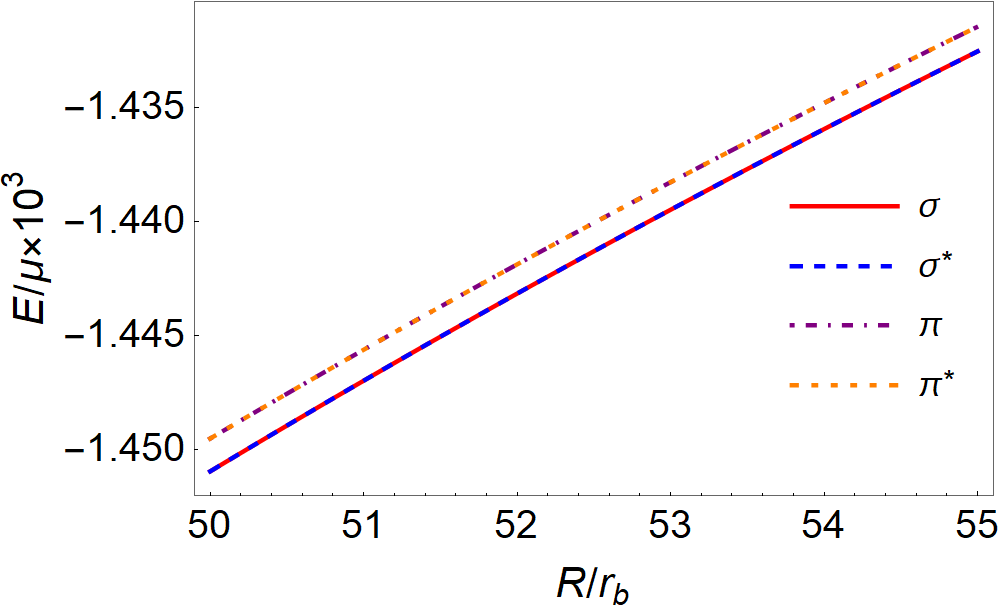}
\caption{Energy eigenvalues of the boson molecular orbits for small orbital separation. Here we choose $q=1$ and $\alpha = 0.1$ as an example.}
\label{fig:energyplot-1}
\end{figure}

\begin{figure}[h]
\includegraphics[width=0.95 \columnwidth]{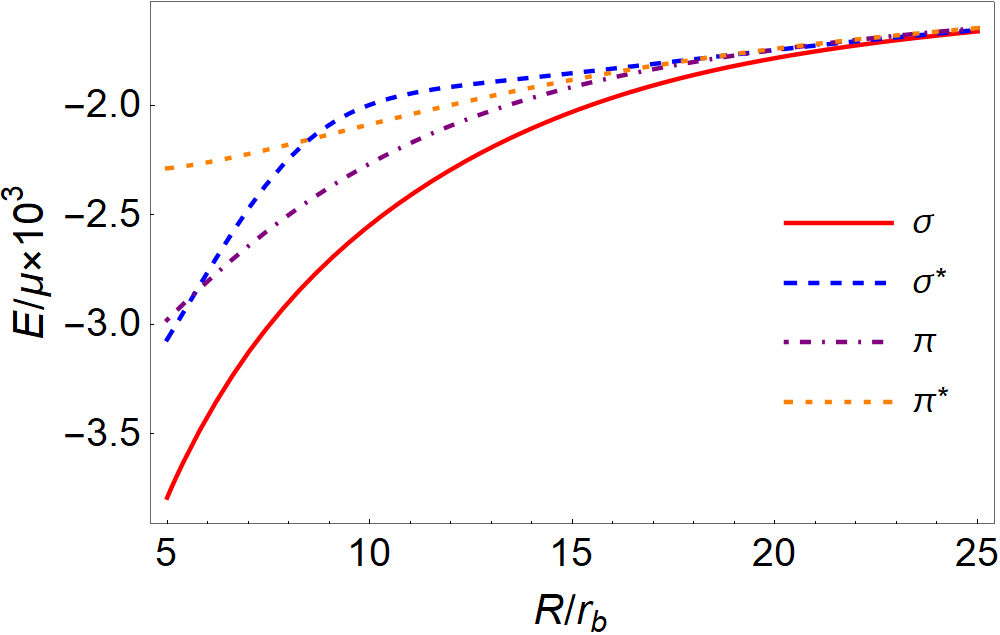}
\caption{Energy eigenvalues of the boson molecular orbits for large orbital separation. Here we choose $q=1$ and $\alpha = 0.1$ as an example. }
\label{fig:energyplot-2}
\end{figure}

We numerically calculate the eigenfrequencies given by Eq.~\eqref{eq:EigenFre} and plot them in Figs.~\ref{fig:energyplot-1} and~\ref{fig:energyplot-2}. It can be seen from Fig.~\ref{fig:energyplot-1} that the four energy levels split and the energy gap is larger when the two black holes are closer. The gap decreases as the orbital separation increases, which is consistent with the fact that the four energy levels are degenerate when $R \rightarrow \infty$. Figure~\ref{fig:energyplot-2} shows that the energy gaps between orbits $\sigma$ and $\sigma^*$, and orbits $\pi$ and $\pi^*$ decrease much faster than that between $\sigma$'s and $\pi$'s orbits. Because the energy gaps between orbits $\sigma$ and $\sigma^*$, and orbits $\pi$ and $\pi^*$ are very close, a question arises as to whether the adiabatic approximation is still valid. 

In the adiabatic approximation, the eigenfunctions and eigenfrequencies vary slowly with time, but the boson remains in the initial energy level during the subsequent evolution, namely, no transition from one energy level to others. In order to see whether the adiabatic approximation is valid for the problem we consider here, we need to check whether the transition between orbits $\sigma$ and $\sigma^*$, and orbits $\pi$ and $\pi^*$ is negligible. 

Consider an arbitrary quantum state 
\begin{eqnarray}
\ket{\psi(t)} = \sum_n c_n(t) \ket{\psi_n(t)},
\end{eqnarray}
where $\ket{\psi_n(t)}$ and $c_n(t)$ are the time-dependent energy eigenstates and their corresponding coefficients, respectively. In the adiabatic approximation, we have $|c_n(t+dt)| = |c_n(t)|$, namely, there is no transition between different energy levels. To estimate the accuracy of the adiabatic approximation, one can check the amplitude of the time derivative of the coefficient $c_n(t)$. It can be shown that (see Appendix~\ref{ap:adiabatic} for details)
\begin{eqnarray}\label{eq:dcdt1}
\frac{d c_k}{d t} = - \sum_{n \neq k} c_n \frac{1}{E_n-E_k} \bra{\psi_k}\frac{\partial \hat{H}}{\partial t}\ket{\psi_n} - i c_k E_k, 
\end{eqnarray}
where the summation characterizes the transition from the initial energy level to other energy levels, and the last term represents the free evolution in the initial energy level.
Now consider the set of state $\{ \ket{\psi_k} \}$ as the four orbits $\ket{\sigma}, \ket{\sigma^*}, \ket{\pi}$ and $\ket{\pi^*}$. Due to the symmetry of the system, it can be shown (see Appendix~\ref{ap:adiabatic} for details) that 
\begin{eqnarray}
\bra{\psi_k}\frac{\partial \hat{H}}{\partial t}\ket{\psi_n} = 0
\label{eq:dhdt0},
\end{eqnarray}
for $n \neq k$. This implies that the adiabatic approximation is a good approximation to the time evolution of the boson. 

Taking into account the emission of gravitational waves and the orbital shrinking, the eigenfrequencies and eigenstates are all time dependent, denoted as $\omega_i(t)$, $\ket{\sigma(t)}, \ket{\sigma^*(t)}, \ket{\pi(t)}$ and $\ket{\pi^*(t)}$, respectively. When the two black holes are infinitely far away, namely, $t \rightarrow -\infty$ and $R \rightarrow \infty$, the normalization constants defined in Eq.~\eqref{eq:Normalization} are the same and equal to one. This implies that the initial state of the boson can be written as 
\begin{eqnarray}
&&\ket{\Phi(-\infty)} \equiv  \ket{\varphi_{2,1,1}^1}  \nonumber\\
&=& \frac{1}{2} \big[ \ket{\sigma(-\infty)} + \ket{\sigma^*(-\infty)} + i \ket{\pi(-\infty)} + i \ket{\pi^*(-\infty)} \big]. \nonumber\\ 
\end{eqnarray}
According to the adiabatic approximation, the subsequent evolution of the state is
\begin{eqnarray}\label{eq:StateEvolution}
&&\ket{\Phi(t)} = \frac{1}{2} \bigg[ e^{-i \int_{-\infty}^t  \omega_1(\tau) \rm{d} \tau} \ket{\sigma(t)} + e^{-i \int_{-\infty}^t  \omega_2(\tau) \rm{d} \tau} \ket{\sigma^*(t)}  \nonumber\\
&& + i \, e^{-i \int_{-\infty}^t  \omega_3(\tau) \rm{d} \tau} \ket{\pi(t)} + i \, e^{-i \int_{-\infty}^t  \omega_4(\tau) \rm{d} \tau} \ket{\pi^*(t)} \bigg],
\end{eqnarray}
We now substitute Eqs.~\eqref{eq:2px} and \eqref{eq:2py} into Eq.~\eqref{eq:StateEvolution} and obtain an expansion of $\ket{\Phi(t)}$ in terms of the isolated orbits $\ket{\varphi_{2,1,\pm 1}^1}$ and $\ket{\varphi_{2,1,\pm 1}^2}$. The coefficient of the state $\ket{\varphi_{2,1, -1}^2}$ is given by
\begin{eqnarray}\label{eq:coefficient2}
&& C_{-}(t) = 
 - \frac{1}{4 N_1} e^{-i \int_{-\infty}^t  \omega_1(\tau) \rm{d} \tau} + \frac{1}{4 N_2} e^{-i \int_{-\infty}^t  \omega_2(\tau) \rm{d} \tau} 
\nonumber\\
&& - \frac{1}{4 N_3 } e^{-i \int_{-\infty}^t  \omega_3(\tau) \rm{d} \tau} + \frac{1}{4 N_4} e^{-i \int_{-\infty}^t  \omega_4(\tau) \rm{d} \tau}.
\end{eqnarray}
The nonzero overlap between the isolated orbit $\ket{\varphi_{2,1, -1}^2}$ and $\ket{\Phi(t)}$ implies that bosons can jump to the orbits of the companion black hole during their evolution. The modular square of $C_{-}(t)$,
\begin{eqnarray}\label{eq:ocpden}
&&|C_{-}(t)|^2 =\sum_{i}\frac{1}{16N_i^2} 
\nonumber\\
&&+ \sum_{i<j} (-1)^{i+j}  \frac{1}{8N_i N_j} \cos \bigg[\int_{-\infty}^t \big(\omega_i(\tau)-\omega_j(\tau) \big) \rm{d} \tau\bigg], \nonumber\\
\end{eqnarray}
represents the probability of jumping to the isolated orbit $\ket{\varphi_{2,1, -1}^2}$. If we only consider the effect of gravitational wave emission and neglect other effects like back reaction of the boson cloud onto the orbital evolution, then we can use the relation defined in Eq.~\eqref{eq:R1} to express the coefficient as a function of $R$, namely, $C_{-}(R)$; and the time integration in Eq.~\eqref{eq:ocpden} can be replaced by an integration over the orbital separation $R$. 

\begin{figure}[h]
\includegraphics[width=0.95 \columnwidth]{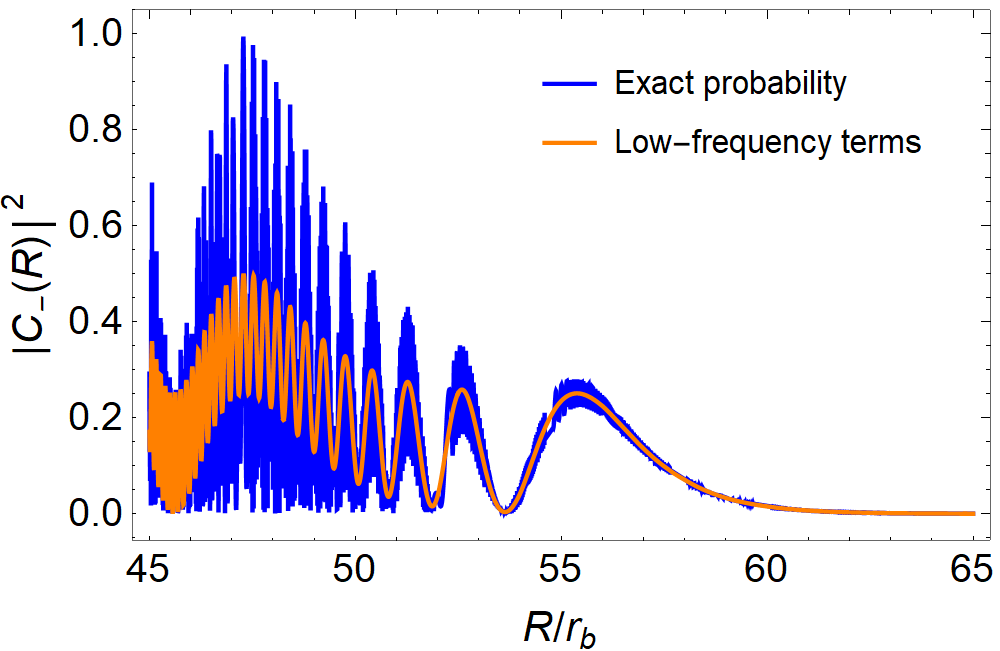}
\caption{ Probability for a boson jumping to the decaying mode $\ket{\psi_{2,1,-1}^2}$ of the companion black hole. Here we choose $q=1$ and $\alpha = 0.1$. The blue curve includes both the rapidly oscillating and slowly oscillating terms, while the orange curve includes only the slowly oscillating terms. }
\label{fig:probplot}
\end{figure}

When $R$ is sufficiently large, the energy levels of orbits $\sigma$ and $\sigma^*$ are almost degenerate ($\omega_1 \approx \omega_2$), as well as that of orbits $\pi$ and $\pi^*$ ($\omega_3 \approx \omega_4$), as can be seen from Fig.~\ref{fig:energyplot-2}. In addition, the normalization factors satisfy $N_1(R) \approx N_2(R)$ and $N_3(R) \approx N_4(R)$. This implies that the coefficient $C_{-}(R)$ approaches to zero when the two black holes are sufficiently far away from each other, which is consistent with the fact that the boson cannot escape to a companion that is very far away. When the two black holes are closer, the energy levels of orbits $\sigma$ and $\sigma^*$ become nondegenerate ($\omega_1 \neq \omega_2$), as well as those of orbits $\pi$ and $\pi^*$. The coefficient $C_{-}(R)$  gradually becomes nonzero, indicating that the boson can transfer to the companion black hole. From Eq.~\eqref{eq:ocpden} we can see that the probability $|C_{-}(R)|^2$ is the sum of six oscillating terms having  frequencies $|\omega_i - \omega_j|$, with $i, j = \{\sigma, \sigma^*, \pi, \pi^* \}$. The frequencies $|\omega_i - \omega_j|$ with $ i \in \{ \sigma, \sigma^* \}$ and $ j \in \{ \pi, \pi^* \}$, or $ j \in \{ \sigma, \sigma^* \}$ and $ i \in \{ \pi, \pi^* \}$, are much higher than that when $ i, j \in \{ \sigma, \sigma^* \}$ or $ i, j \in \{ \pi, \pi^* \}$. 

Figure~\ref{fig:probplot} shows the occupation probability $|C_{-}(R)|^2$ of the orbit $\ket{\psi_{2,1,-1}^2}$, which is one of the decaying modes of the companion black hole. In the numerical calculation, we use a sufficiently large distance $R=80 \,     r_b$ as a replacement for the infinitely large distance ($t = -\infty$) in the lower limit of the integration in Eq.~\eqref{eq:ocpden}, which introduces a negligible error. The blue curve takes into account the full expression of $|C_{-}(R)|^2$ and includes highly oscillating terms. If we remove the highly oscillating terms and only keep two terms with oscillating frequencies $|\omega_{\pi}-\omega_{\pi^*}|$ and $|\omega_{\sigma}-\omega_{\sigma^*}|$, we obtain the profile of the probability $|C_{-}(R)|^2$, which is the orange curve shown in Fig.~\ref{fig:probplot}. The profile captures the change of the probability that the boson occupies the decaying mode of the companion black hole during the evolution of the binary black holes. When $R > 63 \, r_b$, the probability is almost zero. It gradually increases as $R$ decreases and reaches its first maximum around $R = 55.5 \, r_b$. The first maximum of the probability is about $0.3$, showing that a significant amount of boson transfers to the companion black hole and occupies the decaying mode $\ket{\psi_{2,1,-1}^2}$. As the orbital separation $R$ further decreases, the profile of the probability oscillates with a higher and higher frequency. Note that the orbital separation corresponding to the first maximum is larger than the Loche limit, which is about $10 \, r_b$ for $q=1$.


The boson occupying the decaying mode may decay into the black hole, resulting in the depletion of the boson cloud. The boson that transfers to the decaying mode of the companion therefore may decay into the companion black hole. This is another channel other than the hyperfine mixing and Bohr mixing that could result in cloud depletion. Assuming that the back reaction is small and the decay rate of the black hole is not affected by the presence of another black hole, the time evolution of the cloud mass can be described as
\begin{eqnarray}
\frac{d M_c}{d t} = 2 \sum_{i=1}^2 \Gamma_{2,1,-1} |C^i_{2,1,-1}|^2 M_c,
\end{eqnarray}
where $M_c$ is the mass of the boson cloud and $\Gamma_{2,1,-1}$, defined by Eq.~\eqref{eq:decayrate}, is the decay rate of the decaying mode $\ket{\psi_{2,1,-1}}$. Here $|C^1_{2,1,-1}|^2$ is the occupation probability of the decaying mode $\ket{\psi_{2,1,-1}^1}$ of the central black hole and $|C^2_{2,1,-1}|^2$ is the occupation probability of the decaying mode $\ket{\psi_{2,1,-1}^2}$ of the companion black hole, namely, $|C^2_{2,1,-1}|^2 = |C_-(R)|^2$. The summation takes into account the decay of the boson into both the central and companion black holes. 

\begin{figure}[h]
\includegraphics[width=0.95 \columnwidth]{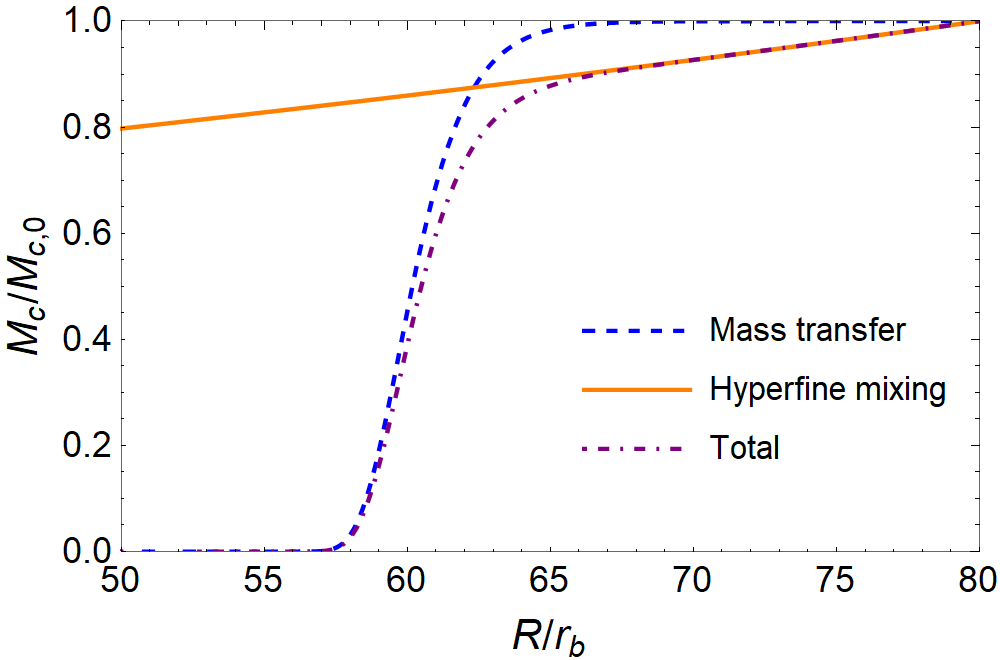}
\caption{Evolution of the boson cloud mass for $\alpha = 0.1$. The blue dashed curve denotes the cloud depletion due to the mass transfer to the decaying mode of the companion black hole, the orange curve denotes the cloud depletion due to the hyperfine mixing of the boson in the central black hole and the purple dashed-dotted curve denotes the total cloud depletion.}
\label{fig:massevoplot}
\end{figure}

To see the effect of mass transfer onto the cloud depletion, we calculate the time evolution of the cloud mass under the assumption that the bosons only decay into the companion black hole. Suppose the total mass of the cloud right before the boson escapes to the companion, e.g., $R = 80 \, r_b$, is $M_{c,0}$. We further assume that the spin of the two black holes are both $\chi = \frac{4\alpha}{1+4\alpha^2}$, which is the critical spin for $\ket{\varphi_{2,1,1}}$ to be saturated~\cite{PhysRevD.99.044001}; and the companion black hole has no bosons around it (Otherwise, we have to modify the initial condition of the state, which in principle can also be handled in our framework). Back reaction of the boson cloud onto the black holes are neglected. The result is shown as the blue dashed curve in Fig.~\ref{fig:massevoplot}. We can see that the boson cloud quickly decays around $R = 60\, r_b$ and finally almost all bosons are absorbed by the companion black hole. To be more specific, the cloud remains unchanged when $R \gtrsim 67 \, r_b$ and almost completely disappears when $R \lesssim 57 \, r_b$. This means all bosons decay into the companion black hole when the probability $|C_{-}(R)|^2$ is approaching to its first maximum. At first glance this seems impossible because the maximal probability is about $0.3$, which is smaller than one. From the physical perspective this is reasonable. Each boson moves between the central and companion black holes. It may be absorbed by the companion black hole when it occupies the orbit $\ket{\varphi_{2,1,-1}}$ of the companion. Though the probability of being absorbed within one round trip is quite small, the boson can travel many round trips when the orbital separation decreases from $67 \, r_b$ to $57 \, r_b$. Therefore, with a very high probability, the boson has been absorbed by the companion black hole when $R < 57 \, r_b$. This is also the key difference from the depletion mechanism due to the hyperfine resonance. It is true that at the resonance point, nearly all bosons jump to the decaying mode, however, the bosons stay in the decaying mode only for a short time, so the total mass that absorbed by the black hole may not be large, as demonstrated in Ref.~\cite{PhysRevD.99.044001}. 

We can also include the cloud depletion due to the hyperfine mixing. Instead of directly calculating the coefficients of the orbits $\ket{\varphi_{2,1,\pm 1}^1}$ from Eq.~\eqref{eq:StateEvolution}, we use the method developed in Ref.~\cite{PhysRevD.99.044001} to calculate the probability that the boson jumps to the decaying mode of the central black hole due to the perturbation of the companion black hole. Firstly, the effect of rotation is not included in Eq.~\eqref{eq:StateEvolution}. Secondly, according to the adiabatic theorem, the slow orbital shrinking cannot induce a direct transition from the growing mode $\ket{\varphi_{2,1,1}^1}$ to the decaying mode $\ket{\varphi_{2,1,-1}^1}$ of the central black hole. A boson may jump to the decaying mode $\ket{\varphi_{2,1,-1}^1}$ by first moving to the companion black and then back to the central black hole. However, this is a second order effect and the probability is at the order of $|C_{-}(R)|^4$, which therefore can be neglected. 

For the parameters that we consider in this paper, the hyperfine resonance occurs at the orbital separation $R \approx 660 \, r_b$, which is much larger than the orbital separation where the mass transfer occurs. We therefore expect that the effect of hyperfine mixing is subdominant in the regime $R<80 \, r_b$, which is confirmed by Fig.~\ref{fig:massevoplot}. The orange curve represents the cloud depletion only due to the hyperfine mixing, which decreases rather slowly as compared to that due to the mass transfer to the companion black hole. The reason is that the probability for a boson jumping to the decaying mode of the central black hole is very small in this regime. Combining the effects of the mass transfer and hyperfine mixing together, we obtain the total cloud depletion, which is shown in Fig.~\ref{fig:massevoplot}. When $ R \gtrsim 67 \, r_b$, the depletion is dominated by the hyperfine mixing; and when $ R \lesssim 67 \, r_b$, the depletion is dominated by the mass transfer to the companion black hole and the decay into it. At about $R \sim 57 \, r_b$, the cloud has completely decayed into the two black holes, which is much earlier than the prediction of hyperfine mixing alone. 

\begin{figure}[h]
\includegraphics[width=0.95 \columnwidth]{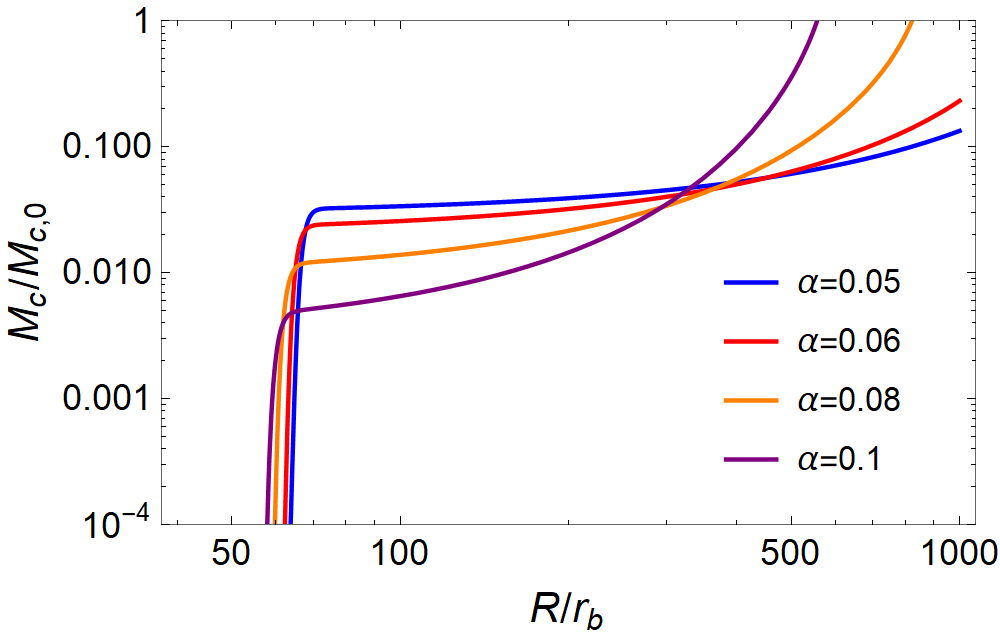}
\caption{Evolution of the boson cloud mass for different $\alpha$ and initial cloud mass $M_{c,0}=\alpha M$~\cite{PhysRevD.96.064050}. }
\label{fig:before}
\end{figure}

In the above discussion we assume that a boson cloud exists when the orbital separation is about $R = 80 \, r_b$. This could happen in several cases. The boson cloud may form when the two black holes are close enough, in particular, when the orbital separation is much smaller than the orbit separation when the hyperfine resonance occurs. In this case the cloud depletes very slowly due to the hyperfine mixing and a significant amount of bosons may still remain when the mass transfer starts to dominate the depletion. The boson cloud may also form when the two black holes are very far away, in particular, when the orbital separation is larger than the orbit separation when the hyperfine resonance occurs. There are two possibilities in this case. If the companion black hole and the cloud are counter-rotating, then the hyperfine resonance never occurs. The hyperfine and Bohr mixing are both weak so the cloud depletes very slowly, as shown in Fig.~\ref{fig:before}, and therefore a substantial amount of cloud may still remain. If the companion black hole and the cloud are co-rotating, then the hyperfine resonance occurs. The hyperfine resonance may cause a strong depletion of the cloud so that only a small amount of cloud is left when the mass transfer starts to dominate the depletion. If this is the case, then the effect of mass transfer does not play a dominant role in the cloud depletion. 



As an example, we consider the case where the companion black hole and the cloud are co-rotating, and the cloud forms after the hyperfine resonance occurs. The cloud has a life time $\tau_c$ that determined only by its own gravitational radiation, which is given by~\cite{yoshino2014gravitational, PhysRevD.96.064050, PhysRevD.99.044001}
\begin{eqnarray}
\tau_c \sim 10^7 \bigg( \frac{M}{3M_\odot} \bigg) \bigg( \frac{0.07}{\alpha} \bigg)^{15} ~ \text{years},
\end{eqnarray}
where $M$ is the mass of the central black hole. We set the initial time of the cloud evolution as the half lifetime of the cloud. The result is shown in Fig.~\ref{fig:before}, where we have included the contribution from the hyperfine mixing, Bohr mixing, mass transfer and gravitational radiation by the cloud itself. For a fixed value of $\alpha$, the cloud depletion due to the hyperfine mixing, Bohr mixing and gravitational radiation dominates at large orbital separation, which proceeds very slowly. When the two black holes are close enough and the mass transfer occurs, the cloud quickly depletes and almost all remaining bosons are absorbed by the companion black hole. For a smaller value of $\alpha$, the cloud depletes slower during the stage of hyperfine mixing and the mass transfer occurs earlier, i.e., at a larger orbital separation. This implies that a larger fraction of boson depletes due to the mass transfer and is absorbed by the companion black hole.  



\section{Conclusions}\label{sec:conclusion}

We develop a framework to study the transfer of bosons between two black holes in a binary black hole system. The framework is formulated through the analogy between the BH-cloud-BH system and the hydrogen molecule ion system in which an electron moves in the potential generated by two protons. When two black holes are sufficiently close, the bosons initially confined around the central black hole can escape to the companion. In the language of quantum mechanics, molecular orbits of the boson form and the boson moves back and forth between two black holes. This results in cloud mass redistribution in the binary black hole system. Furthermore, the boson escapes to the companion may occupy the decaying mode and therefore may decay into the companion black hole. We find that the boson cloud that exists right before the mass transfer occurs completely disappears thgough this mechanism of cloud depletion. The cloud depletion to the companion black hole may have important consequences to the orbital evolution of the binary black hole system and the gravitational waveform emitted from it. 

Our framework is used to study the simplest model where the two black holes have the same mass and spin, and their spin orientation is parallel. It can be straightforwardly generalized to explore more realistic models, for example, the two black holes may have unequal mass and spin, or their spin orientation may be different. This would require a modification to the variational method used to calculate the molecular orbits and the energy eigenvalues. Another interesting case is that the companion may not be a black hole but a compact star. Then there is only cloud mass redistribution but no cloud depletion due to the decay into the companion. However, this could also have important consequences to the evolution of the binary system and their gravitational waveforms.

{\bf Acknowledgements:} D. S. is supported by the Fundamental Research Funds for the Central Universities, HUST (Grant No. 5003012068). Y. M. is supported by the University start-up funding provided by Huazhong University of Science and Technology.


\appendix
\vspace{0.5cm}




\section{Choice of coordinates}

To describe the orbits of the boson in the binary black hole system, we need to set up an appropriate coordinate system, which is schematically shown in Fig.~\ref{fig:configuration}. The origin of the coordinate system is located at the central black hole, and the $z$-axis is parallel to the spin of the central black hole and the $x$-axis is pointing towards the companion black hole. We use spherical coordinates $(r_1, \theta_1, \varphi_1)$ to represent the position of the boson relative to the central black hole, and $(r_2, \theta_2, \varphi_2)$ to denote the position of the boson relative to the companion black hole. Whilst one set of coordinates is sufficient, we introduce the coordinates $(r_2, \theta_2, \varphi_2)$ only for convenience since the isolated wave functions of the boson belonging to the companion black hole can be conveniently expressed using the coordinates $(r_2, \theta_2, \varphi_2)$. The coordinates $(r_2, \theta_2, \varphi_2)$ can be written in terms of the coordinates $(r_1, \theta_1, \varphi_1)$,
\begin{eqnarray}
r_2 &=& \sqrt{r_1^2 + R^2 -2 r_1 R \sin \theta_1 \cos \varphi_1 } ~,
\nonumber\\
\cos \theta_2 &=& \frac{r_1 \cos \theta_1}{r_2}, 
\nonumber\\
\sin \theta_2 &=& \sqrt{1-(\cos \theta_2)^2} ~, 
\nonumber\\
\cos \varphi_2 &=& \frac{r_1 \sin \theta_1 \cos \varphi_1 - R}{r_2 \sin \theta_2}, 
\nonumber\\
\sin \varphi_2 &=& \frac{r_1 \sin \theta_1 \sin \varphi_1}{r_2 \sin \theta_2}.
\end{eqnarray}
By using these relations we can carry out all the calculation using only one set of coordinates, namely, the coordinates $(r_1, \theta_1, \varphi_1)$.





\section{Wave functions and overlap integrals}\label{app:WaveFunc}

The wave functions for the $n=2$  boson around an isolated black hole are given by 
\begin{eqnarray}
\varphi_{2, 1, 1}({\bf r}) &=&  \frac{1}{8}\sqrt{ \frac{1}{ \pi} } r_b^{-5/2} r e^{-r/2 r_b} \sin \theta e^{i \varphi}, 
\nonumber\\
\varphi_{2, 1, -1}({\bf r}) &=&  \frac{1}{8}\sqrt{ \frac{1}{ \pi} } r_b^{-5/2}r e^{-r/2r_b} \sin \theta e^{-i \varphi}, 
\nonumber\\
\varphi_{2p_x}({\bf r}) &=&  \frac{1}{8}\sqrt{ \frac{2}{ \pi} } r_b^{-5/2}r e^{-r/2r_b} \sin \theta \cos \varphi, \nonumber\\
\varphi_{2p_y}({\bf r}) &=&  \frac{1}{8}\sqrt{ \frac{2}{ \pi} } r_b^{-5/2}r e^{-r/2r_b} \sin \theta \sin \varphi, 
\end{eqnarray}
where $r_b$ is the Bohr radius of the boson.

The wave functions for the molecular orbits of the boson in the binary black hole system are given by
\begin{widetext}
\begin{eqnarray}
\varphi_{\sigma}({\bf r}) &=& \frac{1}{8 N_1}\sqrt{ \frac{1}{ \pi} } \, \bigg(r_{b1}^{-5/2}r_1 e^{-r_1/2r_{b1}} \sin \theta_1 \cos \varphi_1 - r_{b2}^{-5/2}r_2 e^{-r_2/2r_{b2}} \sin \theta_2 \cos \varphi_2 \bigg), 
\nonumber\\
\varphi_{\sigma^*}({\bf r}) &=&  \frac{1}{8 N_2}\sqrt{ \frac{1}{ \pi} } \, \bigg(r_{b1}^{-5/2}r_1 e^{-r_1/2r_{b1}} \sin \theta_1 \cos \varphi_1 + r_{b2}^{-5/2}r_2 e^{-r_2/2r_{b2}} \sin \theta_2 \cos \varphi_2 \bigg), 
\nonumber\\
\varphi_{\pi}({\bf r}) &=&  \frac{1}{8 N_3}\sqrt{ \frac{1}{ \pi} } \, \bigg(r_{b1}^{-5/2}r_1 e^{-r_1/2r_{b1}} \sin \theta_1 \sin \varphi_1 + r_{b2}^{-5/2}r_2 e^{-r_2/2r_{b2}} \sin \theta_2 \sin \varphi_2 \bigg), 
\nonumber\\
\varphi_{\pi^*}({\bf r}) &=&  \frac{1}{8 N_4}\sqrt{ \frac{1}{ \pi} } \, \bigg(r_{b1}^{-5/2}r_1 e^{-r_1/2r_{b1}} \sin \theta_1 \sin \varphi_1 - r_{b2}^{-5/2}r_2 e^{-r_2/2r_{b2}} \sin \theta_2 \sin \varphi_2 \bigg),
\end{eqnarray}
where $r_{b1}$ and $r_{b2}$ are the Bohr radius for the boson in central and companion black holes, respectively.
The overlap integrals are given by
\begin{eqnarray}
\langle \varphi_{2p_x}^1 \ket{\varphi_{2p_x}^2} &=& \frac{1}{32 \pi} r_{b1}^{-\frac{5}{2}} r_{b2}^{-\frac{5}{2}} \int r_1^3 r_2 e^{-\frac{r_1}{2r_{b1}}-\frac{r_2}{2r_{b2}}} \sin^2 \theta_1 \sin \theta_2 \cos \varphi_1 \cos \varphi_2 \, {\rm d} r_1 {\rm d} \theta_1 {\rm d} \varphi_1,
\nonumber\\
\bigg \langle \varphi_{2p_x}^1 \bigg| \frac{\alpha}{r_2} \bigg| \varphi_{2p_x}^2 \bigg \rangle &=& \frac{\alpha}{32\pi}r_{b1}^{-\frac{5}{2}}r_{b2}^{-\frac{5}{2}} \int r_1^3 e^{-\frac{r_1}{2r_{b1}}-\frac{r_2}{2r_{b2}}} \sin^2 \theta_1 \sin \theta_2 \cos \varphi_1 \cos \varphi_2 \, {\rm d} r_1 {\rm d} \theta_1 {\rm d} \varphi_1,
\nonumber\\
\bigg\langle \varphi_{2p_x}^1 \bigg| \frac{\alpha}{r_2} \bigg| \varphi_{2p_x}^1 \bigg\rangle &=& \frac{\alpha}{32\pi}r_{b1}^{-5} \int \frac{r_1^4}{r_2} e^{-\frac{r_1}{r_{b1}}} \sin^3 \theta_1 \cos^2 \varphi_1 \, {\rm d} r_1 {\rm d} \theta_1 {\rm d} \varphi_1,
\nonumber\\
\langle \varphi_{2p_y}^1 \ket{\varphi_{2p_y}^2} &=& \frac{1}{32\pi}r_{b1}^{-\frac{5}{2}}r_{b2}^{-\frac{5}{2}} \int r_1^3 r_2 e^{-\frac{r_1}{2r_{b1}}-\frac{r_2}{2r_{b2}}} \sin^2 \theta_1 \sin \theta_2 \sin \varphi_1 \sin \varphi_2 \, {\rm d} r_1 {\rm d} \theta_1 {\rm d} \varphi_1,
\nonumber\\
\bigg \langle \varphi_{2p_y}^1 \bigg| \frac{\alpha}{r_2} \bigg| \varphi_{2p_y}^2 \bigg \rangle &=& \frac{\alpha}{32\pi}r_{b1}^{-\frac{5}{2}}r_{b2}^{-\frac{5}{2}} \int r_1^3 e^{-\frac{r_1}{2r_{b1}}-\frac{r_2}{2r_{b2}}} \sin^2 \theta_1 \sin \theta_2 \sin \varphi_1 \sin \varphi_2 \, {\rm d} r_1 {\rm d} \theta_1 {\rm d} \varphi_1,
\nonumber\\
\bigg \langle \varphi_{2p_y}^1 \bigg| \frac{\alpha}{r_2} \bigg| \varphi_{2p_y}^1 \bigg \rangle &=& \frac{\alpha}{32\pi}r_{b1}^{-5} \int \frac{r_1^4}{r_2} e^{-\frac{r_1}{r_{b1}}} \sin^3 \theta_1 \sin^2 \varphi_1 \, {\rm d} r_1 {\rm d} \theta_1 {\rm d} \varphi_1.
\end{eqnarray}

\end{widetext}

\section{Quantum adiabatic theorem}\label{ap:adiabatic}

In this appendix, we show that the adiabatic approximation can be applied in our case. 
Since the potential generated by the two black holes changes very slowly, we assume that the energy eigenvalues also evolve slowly and continuously with time, and they satisfy the equation
\begin{eqnarray}\label{eq:eigenvalue}
\hat{H}(t) \ket{\psi_n(t)} = E_n(t) \ket{\psi_n(t)}.
\end{eqnarray}
We assume that the energy levels are not degenerate. An arbitrary state at a given time can be written as
\begin{eqnarray}\label{eq:sum}
\ket{\psi(t)} = \sum_{n} c_n(t) \ket{\psi_n(t)},
\end{eqnarray}
and satisfies the Schr\"odinger equation
\begin{eqnarray}\label{eq:SEequation}
i \frac{\partial}{\partial t} \ket{\psi(t)} = \hat{H}(t) \ket{\psi(t)}, 
\end{eqnarray}
By substituting Eq.~\eqref{eq:sum} into Eq.~\eqref{eq:SEequation}, and multiplying both sides from the left by $\bra{\psi_k}$, we have
\begin{eqnarray}
i\frac{\partial c_k}{\partial t}+i\sum_n c_n\langle\psi_k \ket{\frac{\partial\psi_n}{\partial t}} =c_k E_k. 
\end{eqnarray}

Now we are going to derive the expression for $\langle\psi_k\ket{\frac{\partial\psi_n}{\partial t}}$. We start with a time derivation of Eq.~\eqref{eq:eigenvalue},
\begin{eqnarray}
\frac{\partial\hat{H}}{\partial t} \ket{\psi_n} + \hat{H} \ket{ \frac{\partial\psi_n}{\partial t} } = 
\frac{\partial E_n }{\partial t} \ket{\psi_n} + E_n \ket{\frac{\partial\psi_n}{\partial t}}.
\end{eqnarray}
For $k \neq n$, we multiply both sides from the left by $\bra{\psi_k}$ and obtain
\begin{eqnarray}
\bra{\psi_k} \frac{\partial\psi_n}{\partial t}\rangle = \frac{1}{E_n-E_k}\bra{\psi_k} \frac{\partial\hat{H}}{\partial t}\ket{\psi_n}. 
\end{eqnarray}

We also need to consider the case when $k = n$, namely, the expression for $\bra {\psi_k} \frac{\partial\psi_k}{\partial t} \rangle$. Taking time derivative of the normalization condition
\begin{eqnarray}
\bra{\psi_k} \psi_k \rangle=1,  
\end{eqnarray}
we have
\begin{eqnarray}
\bra{\psi_k} \frac{\partial \psi_k}{\partial t} \rangle + \bra{\frac{\partial\psi_k}{\partial t}} \psi_k \rangle=0. 
\end{eqnarray}
It is evident that $\langle \psi_k \ket{\frac{\partial \psi_k}{\partial t}}$ is purely imaginary, which in principle can be canceled by appropriately adding a phase. Therefore, the time derivative of the coefficient $c_n$ is given by 
\begin{eqnarray}
\frac{d c_k}{d t}=-\sum_{n \neq k} c_n \frac{1}{E_n-E_k} \bra{\psi_k} \frac{\partial\hat{H}}{\partial t}\ket{\psi_n} - i c_k E_k. 
\end{eqnarray}

We now show that $\bra{\psi_k}\frac{\partial \hat{H}}{\partial t}\ket{\psi_n} = 0$ when $n \neq k$, which is the equality given by Eq.~\eqref{eq:dhdt0}. Since $n, k \in \{  \sigma, \sigma^*, \pi, \pi^* \}$, so there are six off-diagonal elements for $\frac{\partial \hat{H}}{\partial t}$. Let us first consider $\bra{\sigma}\frac{\partial \hat{H}}{\partial t}\ket{\sigma^*}$ and $\bra{\pi}\frac{\partial \hat{H}}{\partial t}\ket{\pi^*}$. 
\begin{eqnarray}
&&\left \langle \sigma \left| \frac{\partial \hat{H}}{\partial t} \right| \sigma^\ast \right \rangle \nonumber\\
&=&\left \langle \varphi_{2p_x}^1 \left| \frac{\partial\hat{H}}{\partial t} \right| \varphi_{2p_x}^1 \right \rangle -\left \langle \varphi_{2p_x}^2 \left| \frac{\partial\hat{H}}{\partial t} \right| \varphi_{2p_x}^2 \right \rangle \nonumber\\
&+&\left \langle \varphi_{2p_x}^1 \left| \frac{\partial\hat{H}}{\partial t} \right| \varphi_{2p_x}^2 \right \rangle-\left \langle \varphi_{2p_x}^2 \left| \frac{\partial\hat{H}}{\partial t} \right| \varphi_{2p_x}^1 \right \rangle. 
\end{eqnarray}
In our simple model we assume that two black holes have exactly the same parameters, so the isolated orbits of the boson are the same for the central and companion black holes. Due to the symmetry of the configuration of the BH-cloud-BH system, we have $\bra{ \varphi_{2p_x}^1}
\frac{\partial\hat{H}}{\partial t}
\ket{\varphi_{2p_x}^1} = \bra{ \varphi_{2p_x}^2}
\frac{\partial\hat{H}}{\partial t}
\ket{\varphi_{2p_x}^2}$. The wave functions $\varphi_{2p_x}^i$ are real and the time derivative of the Hamiltonian is Hermitian, so we have $\bra{ \varphi_{2p_x}^1}
\frac{\partial\hat{H}}{\partial t}
\ket{\varphi_{2p_x}^2} = \bra{ \varphi_{2p_x}^2}
\frac{\partial\hat{H}}{\partial t}
\ket{\varphi_{2p_x}^1}$. Therefore, we find that $\bra{\sigma}\frac{\partial \hat{H}}{\partial t}\ket{\sigma^*}=0$. By using similar arguments we also have $\bra{\pi}\frac{\partial \hat{H}}{\partial t}\ket{\pi^*} = 0$.


To calculate other elements of $\frac{\partial \hat{H}}{\partial t}$, we need to know its explicit expression, which is given by 
\begin{eqnarray}
\frac{\partial \hat{H}}{\partial t} 
&=& \frac{\partial \hat{H}}{\partial r_2} \frac{\partial r_2} {\partial R} 
\frac{\partial R}{\partial t} =\frac{\alpha}{r_2^2}\frac{\partial r_2}{\partial R}\frac{\partial R}{\partial t}
\nonumber \\
&=& - \frac{64}{5} \alpha \, q(1+q)  \frac{M^3}{R^3} \frac{R-r_1 \sin \theta_1 \cos\varphi_1}{r_2^3}.
\end{eqnarray}
By using the explicit expressions of the orbits given by Eqs.~\eqref{eq:2px} and \eqref{eq:2py}, we find that other elements of $\frac{\partial \hat{H}}{\partial t}$ can be expanded using $\bra{\varphi_{2p_x}^i} \frac{\partial \hat{H}}{\partial t}\ket{\varphi_{2p_y}^j}$, with $i, j \in \{1, 2 \}$. We now show that  $\bra{\varphi_{2p_x}^i} \frac{\partial \hat{H}}{\partial t}\ket{\varphi_{2p_y}^j} = 0$ for all choices of $i, j$. 


Neglecting the constant factor that is independent of $(r_1, \theta_1, \varphi_1)$, we find
\begin{eqnarray}
&&\left\langle \varphi_{2p_x}^1 \left| \frac{\partial\hat{H}}{\partial t} \right| \varphi_{2p_y}^1 \right\rangle 
\nonumber\\
&\sim& \int {\rm d} V \, \frac{R - r_1 \sin \theta_1 \cos \varphi_1}{r_2^3} r_1^2 e^{-r_1/r_b} \sin^2 \theta_1 \sin \varphi_1 \cos \varphi_1 
\nonumber\\
&\sim& \int {\rm d} x {\rm d} y {\rm d} z \, \frac{x(R - x) y}{(r_1^2 + R^2 - 2 R x)^{3/2}} \, e^{-r_1/r_b},
\end{eqnarray}
where we have used the Cartesian coordinates $(x, y, z)$, 
\begin{eqnarray}
x = r_1 \sin \theta_1 \cos \varphi_1, ~~ 
y = r_1 \sin \theta_1 \sin \varphi_1, ~~
z = r_1 \cos \theta_1.  \nonumber
\end{eqnarray}
It is evident that the integrand is an odd function of $y$, so the integral is zero. 


\begin{eqnarray}
&&\left\langle \varphi_{2p_x}^1 \left| \frac{\partial\hat{H}}{\partial t} \right| \varphi_{2p_y}^2 \right\rangle 
\nonumber\\
&\sim&\int {\rm d} V \frac{R - r_1 \sin \theta_1 \cos \varphi_1}{r_2^3} e^{-\left(r_1+r_2\right)/2 r_b} r_1 \sin \theta_1 \cos \varphi_1 \nonumber\\
&&
\times \, r_2 \sin \theta_2 \sin \varphi_2
\nonumber\\
&\sim& \int {\rm d} x {\rm d} y {\rm d} z \, \frac{x(R - x) y}{(r_1^2 + R^2 - 2 R x)^{3/2}}
\, e^{-(r_1 + r2)/2 r_b}, 
\end{eqnarray}
where $r_2 = \sqrt{r_1^2 + R^2 - 2 R x}$, and we have used the equality $ y = r_1 \sin \theta_1 \cos \varphi_1 = r_2 \sin \theta_2 \cos \phi_2$. It is evident that the integrand is an odd function of $y$, so the integral is zero.


By using the symmetry of the BH-cloud-BH system, we have
\begin{eqnarray}
\left \langle \varphi_{2p_x}^2 \left| \frac{\partial\hat{H}}{\partial t} \right| \varphi_{2p_y}^1 \right \rangle
&=&
- \left \langle \varphi_{2p_x}^1 \left| \frac{\partial\hat{H}}{\partial t} \right| \varphi_{2p_y}^2\right\rangle = 0,
\nonumber\\
\left \langle \varphi_{2p_x}^2 \left| \frac{\partial\hat{H}}{\partial t} \right| \varphi_{2p_y}^2 \right \rangle
&=&
- \left \langle \varphi_{2p_x}^1 \left| \frac{\partial\hat{H}}{\partial t} \right| \varphi_{2p_y}^1\right\rangle = 0.
\nonumber\\
\end{eqnarray}
Therefore, We have  $\bra{\varphi_{2p_x}^i} \frac{\partial \hat{H}}{\partial t}\ket{\varphi_{2p_y}^j} = 0$ for all choices of $i, j$. As a result, all other elements of $\frac{\partial \hat{H}}{\partial t}$ are zero. In summary, we have
\begin{eqnarray}
\left \langle \psi_k \left| \frac{\partial \hat{H}}{\partial t} \right| \psi_n \right \rangle = 0
\end{eqnarray}
for $n \neq k$. 


\bibliography{ref_BosonCloud}

\end{document}